# Interfacial Spin-Orbit Coupling Induced Room-Temperature Ferromagnetic Insulator

Yuhao Hong,[1,2] Shilin Hu,[1] Ziyue Shen,[1] Chao Deng,[3] Xiaodong Zhang,[3] Lei Wang,[1] Long Wei,[1] Qinghua Zhang,[4] Lingfei Wang,[5] Liang Si,[3,6,7,‡] Yulin Gan,[1,#] Kai Chen,[1,†] Zhaoliang Liao[1,*]

[1]National Synchrotron Radiation Laboratory, University of Science and Technology of China, Hefei 230029, China.
[2]MESA+ Institute for Nanotechnology, University of Twente, Enschede 7500 AE, the Netherlands.
[3]School of Physics, Northwest University, Xi'an 710127, China
[4]Beijing National Laboratory for Condensed Matter Physics, Institute of Physics, Chinese Academy of Sciences, Beijing 100190, China.
[5]National Research Center for Physical Sciences at Microscale, University of Science and Technology of China, Hefei 230026, China.
[6]Shaanxi Key Laboratory for Theoretical Physics Frontiers, Xi'an 710127, China
[7]Institute of Solid State Physics, TU Wien, 1040 Vienna, Austria

**ABSTRACT**. Fabricating room-temperature ferromagnetic insulators, which are crucial candidates for next-generation dissipation-free quantum and spintronic devices, remains a significant challenge. In this Letter, we report on the epitaxial synthesis of novel room-temperature ferromagnetic insulating thin films created through the precise construction of (111)-oriented 3d/5d interfaces. Our analysis indicates that, unlike conventional doping methods, the (111)-oriented $SrIrO_3/La_{2/3}Sr_{1/3}MnO_3$ (SIO/LSMO) interfaces exhibit markedly enhanced spin-orbit coupling. This enhanced interfacial spin-orbit coupling strengthens the electron-phonon coupling in LSMO, thereby shortening the electronic mean free path. As a result, the intrinsic metallicity of LSMO is suppressed, giving rise to a new ferromagnetic insulating phase that emerges between the ferromagnetic metal and paramagnetic insulator regimes of the LSMO phase diagram. Furthermore, the temperature window of the ferromagnetic insulating phase can be tuned by precisely controlling the thickness of the LSMO layers. Our Letter reveals a new strategy for developing ferromagnetic insulators by engineering 3d/5d interfaces and orientations, paving a way for the development of novel dissipation-free quantum and spintronic devices.

## I. INTRODUCTION.

Room-temperature ferromagnetic insulators (FMIs) are essential for the development of next-generation dissipation-free spintronic [1-5] and quantum [6-8] devices due to their ability to block electric currents, enabling the generation of pure spin currents or the manipulation of spins in non-magnetic layers via magnetic proximity effects [9,10]. However, ferromagnetism in nature typically arises from double-exchange interactions [11] or Ruderman-Kittel-Kasuya-Yosida [12-14] interactions, both of which involve itinerant electrons and thus imply metallicity, complicating the artificial construction of FMIs. Most reported oxide FMIs exhibit Curie temperatures far below room temperature, and often near or below liquid nitrogen temperature [14-16], severely limiting their practical applications. Consequently, room-temperature FMIs in oxides are exceedingly rare, with only a few reports achieving near-room-temperature FMIs through approaches such as the octahedral proximity effect [17], inducing oxygen vacancies [18], or using double perovskites [19].

$La_{2/3}Sr_{1/3}MnO_3$ (LSMO), a classic perovskite material, is considered an ideal candidate for spintronic applications due to its rich array of physical properties, including colossal magnetoresistance (CMR) [20,21], metal-insulator transitions [22], spin textures [23], and oxygen octahedral rotations [24,25], as well as its excellent room-temperature magnetoelectric properties and cost-effectiveness [26]. However, the ferromagnetism in LSMO originates from the double exchange interaction, where electrons hop between $Mn^{3+}$ and $Mn^{4+}$ ions via oxygen bridges [11]. This coupling of metallic behavior with ferromagnetism leads to increased power consumption and presents a fundamental obstacle to the development of FMIs in LSMO.

Unlike the (100) and (110) planes, the perovskite (111) plane exhibits a graphene-like honeycomb lattice between layers [27-29]. The unique symmetry breaking in this plane, combined with strong spin-orbit coupling (SOC) [30], is predicted to give rise to various novel quantum [31] and topological [32,33] effects. The iridate family, with its low-spin $d^5$ configuration of $Ir^{4+}$ in an octahedral crystal field that leads to half-filled SOC double peaks ($J_{eff}$ = 1/2 state) [34,35], represents a promising candidate for exploring SOC effects, giving rise to unconventional spin-orbit states that have been both theoretically predicted [36] and experimentally observed [37]. However, the perovskite phase of $SrIrO_3$ (SIO), which possesses strong SOC, is metastable and tends to become monoclinic [28,38] when it is grown on (111)-oriented perovskite substrates, posing significant challenges for sample construction.

In this Letter, we demonstrate the realization of ultra-thin room-temperature FMIs by the rational design of (111)-oriented SIO/LSMO superlattices, engineered through precise control of orientation and interfaces. Remarkably, the thickness of LSMO in a single superlattice period is as low as ~1.1 nm, yet the Curie temperature ($T_c$) of LSMO surprisingly exceeds 296 K. X-ray magnetic circular dichroism (XMCD) analysis confirms the presence of significant ferromagnetic (FM) signals of Mn at 300 K in the ferromagnetic insulating sample. Furthermore, magnetoconductance (MC) analysis reveals strong SOC persisting near to room temperature. The presence of strong interfacial SOC appears to influence polarons in LSMO, offering a plausible route for increased electron-phonon coupling [39,40] and suppressing the intrinsic metallic state. Density-functional theory (DFT) calculations further show that interfacial charge transfer is strongly suppressed at the (111)-oriented LSMO/SIO interface, excluding direct electronic interactions across the interface. These results are consistent with the notion that interfacial SOC plays an important role in stabilizing the insulating state in these ultra-thin LSMO films.

## II. METHODS

A series of $(SIO_m/LSMO_n)_{10}$ superlattices [$(I_mM_n)_{10}$, $1 \leq m \leq 4$, $1 \leq n \leq 15$] was grown on atomically smooth Ti-terminated $SrTiO_3$ (STO) (111) substrates [Fig. S1(a)] by our home-made pulsed laser deposition system, where the reflection high-energy electron diffraction is employed to precisely control the epitaxial growth of the thin films. All electrical transport measurements were conducted using the Van der Pauw method following the completion of magneto tests. More details regarding superlattices growth, sample characterization, and DFT calculations can be found in the Supplemental Material [41].

## III. RESULTS

The structural quality of the superlattices is shown in Fig. 1(a). A cross-sectional view of the (-1 -1 2)-oriented $(I_2M_{15})_{10}$ sample [Fig. 1(b)] obtained by HAADF-STEM [Fig. 1(c)] reveals atomically sharp interfaces with minimal interdiffusion, consistent with large-scale STEM (Fig. S2). Additional XRR (Fig. S3) and reciprocal space mapping (Fig. S4) measurements corroborate the high-energy electron diffraction observations. These results indicate that the in-plane biaxial compressive stress from the substrate and LSMO layers stabilizes the perovskite phase of SIO[28,38,54].

Figure 2(a) presents the temperature-dependent resistivity for $(I_2M_n)_{10}$ superlattices with varying LSMO thicknesses. For LSMO thicknesses $n$ in a single superlattice period below 4 unit cells (thickness $t \leq 0.88$ nm), the superlattices exhibit a typical insulating behavior, with resistivity at ~75 K exceeding the range of the measurement electronics. When $n$ increases to 5 ($t$ ~1.1 nm), a metal-insulator transition (MIT) appears around 150 K. As $n$ is further increased, the MIT temperature ($T_{MIT}$) progressively shifts to room temperature, which is consistent with the "dead layer" behavior [22,55] observed in the (001)-oriented LSMO films. Figure 2(b) shows the angle dependence of magnetoresistance (MR) at 9 T across full temperature ranges, where the magnetic field is rotated out-of-plane with respect to the current direction. The angle-resolved MR indicates that the easy axis of magnetization lies in-plane, and the magnetic anisotropy is uniaxial with π-periodic twofold rotational symmetry. This observation aligns well with our expectations based on shape anisotropy [56].

The out-of-plane and in-plane MR traces [Figs. 2(c) and 2(d)] display the characteristic CMR response of LSMO. The butterfly-like hysteresis in the out-of-plane geometry collapses upon approaching $T_{MIT}$ [inset of Fig. 2(c)], coincident with the reduction of the coercive field observed in the in-plane configuration [inset of Fig. 2(d)]. Below $T_{MIT}$ (~150 K), the pronounced low-field cusps in the in-plane MR provide additional evidence that the magnetic easy axis lies in the film plane.

*Contact author: zliao@ustc.edu.cn
†Contact author: kaichen2021@ustc.edu.cn
#Contact author: ylgan@ustc.edu.cn
‡Contact author: siliang@nwu.edu.cn

Interestingly, the magnetic properties show distinct behavior compared to the electrical transport properties. As shown by the magnetic moment-temperature curves in Fig. 3(a), the $T_c$ begins to emerge at $n = 3$, even as electrical transport retains insulating behavior, indicating that the magnetic dead layer is thinner than the electrical dead layer. At $n = 5$, $T_c$ already approaches room temperature (~296 K). The phase diagram in Fig. 3(b) clearly reveals the emergence of a novel ferromagnetic insulating state during the transition from the ferromagnetic metal to the paramagnetic insulator state. The FMI phase extends up to around $n = 15$, allowing for a tunable temperature range for the phase by adjusting the LSMO thickness, with the widest temperature range (from 150 to 296 K) of the phase achieved at $n = 5$. Figure 3(c) illustrates the temperature dependence of magnetization versus field for the $I_2M_5$ sample, showing square-like FM hysteresis loops below $T_{MIT}$ (~150 K). The coercive force diminishes above $T_{MIT}$, with the sample maintaining weak ferromagnetism (~0.1 $\mu_B$/Mn) at 300 K, which further confirms the room-temperature ferromagnetism. In addition, the disparity between the in-plane and out-of-plane magnetization-field curves at 10 K [Fig. 3(d)] further supports the conclusion that the easy axis of magnetization lies in-plane, consistent with the observed electrical transport behavior.

## IV. DISCUSSION

To further elucidate the origin of the room-temperature FM state, we investigated the effect of varying the SIO thickness $m$. If SIO were responsible for the FM, an increase in $m$ should enhance the FM signal. However, neither $T_c$ [Fig. S5(a)] nor the saturation magnetization [Fig. S5(b)] show any notable change with $m$, excluding SIO as the primary magnetic contributor. Meanwhile, as $m$ increases from 2 to 3, the magnetic easy axis undergoes a transition from in-plane to out-of-plane [Fig. S5(b)], consistent with the previous study [35]. To further exclude the possibility of impurity interference, Mn $L_{2,3}$-edge XMCD analysis was conducted to directly confirm the intrinsic FM of LSMO. As shown in Fig. 3(a) and S6(a)-(c), strong non-zero XMCD signals of the $I_2M_5$ sample were detected at 4 K, 200 K and 300 K, providing compelling evidence that the observed ferromagnetism in these ultrathin films originates from Mn rather than impurities. Notably, a distinct peak at ~640 eV was detected in the circularly polarized x-ray absorption spectroscopy, while no significant signal was observed at ~639 eV. This peak corresponds to $Mn^{3+}$ under substantial compressive stress near the surface [57,58], rather than $Mn^{2+}$, suggesting that the room-temperature ferromagnetism is driven by double exchange interactions between $Mn^{3+}$ and $Mn^{4+}$ ions in LSMO.

While prior studies have suggested that SIO can enhance the FM of (001)- and (110)-oriented LSMO and that this enhancement is often accompanied by increased metallicity [35] attributed to the double-exchange interaction. To further explore the origin of the FM enhancement in our samples, we substituted SIO in the superlattice [Fig. S5(a)] with STO. The $T_c$ of the $(STO_2/LSMO_5)_{10}$ ($T_2M_5$) sample (~150 K) closely aligned with its $T_{MIT}$, indicating that the enhanced FM in LSMO within the superlattice is primarily induced by SIO rather than being dominated by orientation.

Although we engineer a robust room-temperature ferromagnetic insulating state, FM in LSMO is typically linked to double exchange, making the insulating behavior nontrivial. To understand the ferromagnetic insulating phase, we performed DFT PDOS charge calculations for (001)- and (111)-oriented LSMO:SIO (3:3) superlattices [Figs. 4(a) and 4(c)]. The structures preserve ideal stoichiometry without oxygen vacancies. Orbital occupations and spin-resolved density of states reveal that charge transfer is strongly suppressed at the (111) interface, whereas in (001) superlattices the Mn-O-Ir bonds enhance hybridization between Mn-$3d_{yz}/3d_{xz}$ orbitals and Ir-$5d_{yz}/5d_{xz}$ orbitals [59]. Meanwhile, the density of states [Fig. 4(b) and 4(d)] and magnetic moments (Table S3) from DFT show that in the (111) orientation, Ir ions exhibit markedly enhanced spin splitting and larger magnetic moments at ground state, consistent with our experimental observations (Fig.S5). Furthermore, previous reports [35] and the annular bright-field image results shown in Fig. S7 confirm that SIO does not suppress the oxygen octahedral rotations in LSMO, implying that strong interfacial SOC could be crucial in inducing the ferromagnetic insulating phase.

Factors such as oxygen vacancies [60], oxygen octahedral rotations [24], disorder [22], and polarons [61,62] have been extensively studied for their influence on the electrical transport properties of LSMO. Among these, the first three factors also affect ferromagnetism, whereas polarons do not alter magnetic transport behavior but significantly influence the MIT. Therefore, a plausible hypothesis is that LSMO generally exhibits large polarons due to strong electron-phonon coupling. Concurrently, the highly symmetry-broken, graphene-like honeycomb lattice within the (111) plane significantly enhances the SOC [63] at the interface between SIO and LSMO. A possible interplay between strong interfacial SOC and large polarons may further enhance electron-phonon coupling, thereby reducing the mean free path of carriers in LSMO. This prevents the formation of an extended conduction network, localizes carriers within small regions, and generates robust gap states that hinder the MIT [39,40]. These effects account for the emergence of the ferromagnetic insulating phase observed in transport.

*Contact author: zliao@ustc.edu.cn
†Contact author: kaichen2021@ustc.edu.cn
#Contact author: ylgan@ustc.edu.cn
‡Contact author: siliang@nwu.edu.cn

Assessing the electron mean free path through theoretical models in the presence of polarons and interfacial SOC is challenging due to their complex interplay. However, the weak antilocalization (WAL) phenomenon can provide a powerful method for probing the strength of interfacial SOC [64,65]. This is typically achieved by measuring the magnetoconductance under an out-of-plane magnetic field to determine the interfacial SOC intensity. To demonstrate such an approach, we provide a detailed analysis in the End Matter. The results show a strong SOC at the superlattice interface near room temperature.

Competition between FM and interfacial SOC has been predicted [66], and observed in dilute magnetic semiconductors [67]. One possible explanation from Dugaev theory [66] states that strong FM exchange produces spin splitting comparable to or larger than the Landau-level spacing, yielding weak localization (WL) or negative MR and suppressing WAL. Thus, robust FM suppresses SOC effects. This scenario matches our observations: at low temperature, double-exchange FM in LSMO dominates and supports metallicity; at higher temperature, weakened FM allows SOC-polaron coupling to dominate, suppressing the MIT and stabilizing the ferromagnetic insulating state. Increasing LSMO thickness strengthens intrinsic FM, thereby reducing SOC influence, shifting $T_{MIT}$ upward, and narrowing the ferromagnetic insulating window. This explains why the ferromagnetic insulating phase with the broadest temperature range appears near the metallic dead-layer thickness. Notably, the higher in-plane symmetry of the (001) superlattices yields weaker SOC ($L_{SO(001)} > L_{SO(111)}$) than in the (111) orientation, further suppressing WAL. Charge transfer at the (001) interface also enhances conduction paths and metallicity. Consequently, (001)-oriented superlattices do not host an ferromagnetic insulating phase.

## V. CONCLUSIONS

In conclusion, we have successfully fabricated a new high-quality $(I_mM_n)_{10}$ superlattice structure on (111)-oriented STO substrates. The graphene-like honeycomb lattice within the perovskite (111) plane significantly enhances the interfacial SOC. Based on MC analysis, we propose that the electrical transport behavior of the superlattice samples is dominated by the FM double-exchange interaction in LSMO below a critical temperature. However, above this critical temperature, WAL emerges at the interface and becomes dominant. Our results point to a possible scenario where strong interfacial SOC might interact with large polarons in LSMO, leading to enhanced electron-phonon coupling and a reduced electronic mean free path, suppressing the intrinsic MIT to favor the emergence of the ferromagnetic insulating phase. Enhancing the interfacial SOC through the (111)-oriented surface to create room-temperature FMIs presents a promising avenue for the development of novel dissipation-free quantum and spintronic devices.

## ACKNOWLEDGMENTS

We thank Guus Rijnders and Gertjan Koster for invaluable discussion. This work was supported by the National Key R&D Program of China (No. 2022YFA1403000) and the National Natural Science Foundation of China (No. 52272095, No. 1974325, No. 12204523, No. 12275272, and No. 12422407). This work was partially carried out at the Deimos beamline at synchrotron Soleil; XMCD-a beamline at the Hefei Light Source; Instruments Center for Physical Science, University of Science and Technology of China; the MESA+ Institute for Nanotechnology; and the USTC Center for Micro and Nanoscale Research and Fabrication.

## DATA AVAILABILITY

The data that support the findings of this article are not publicly available upon publication because it is not technically feasible and/or the cost of preparing, depositing, and hosting the data would be prohibitive within the terms of this research project. The data are available from the authors upon reasonable request.

*Contact author: zliao@ustc.edu.cn
†Contact author: kaichen2021@ustc.edu.cn
#Contact author: ylgan@ustc.edu.cn
‡Contact author: siliang@nwu.edu.cn

*Contact author: zliao@ustc.edu.cn
†Contact author: kaichen2021@ustc.edu.cn
#Contact author: ylgan@ustc.edu.cn
‡Contact author: siliang@nwu.edu.cn

*Contact author: zliao@ustc.edu.cn
†Contact author: kaichen2021@ustc.edu.cn
#Contact author: ylgan@ustc.edu.cn
‡Contact author: siliang@nwu.edu.cn

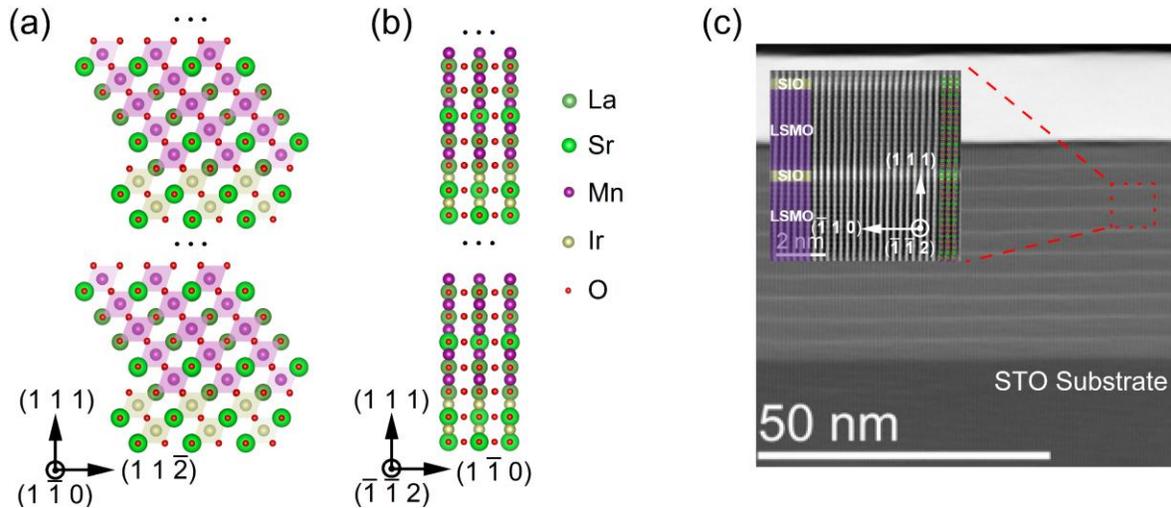

**FIG 1. Superlattice structure characterization.** (a) Schematic cross-sectional view of the superlattice along the (1 -1 0) orientation, the purple quadrilateral represents the oxygen octahedron. (b) Schematic cross-sectional view of the superlattice along the (-1 -1 2) orientation. (c) HAADF-STEM image of the $I_2M_{15}$ superlattice sample. The illustration is a zoom-in image of the red dashed box area. The right side of the illustration shows the atomic structure of the (-1 -1 2)-oriented cross section.

*Contact author: zliao@ustc.edu.cn
†Contact author: kaichen2021@ustc.edu.cn
#Contact author: ylgan@ustc.edu.cn
‡Contact author: siliang@nwu.edu.cn

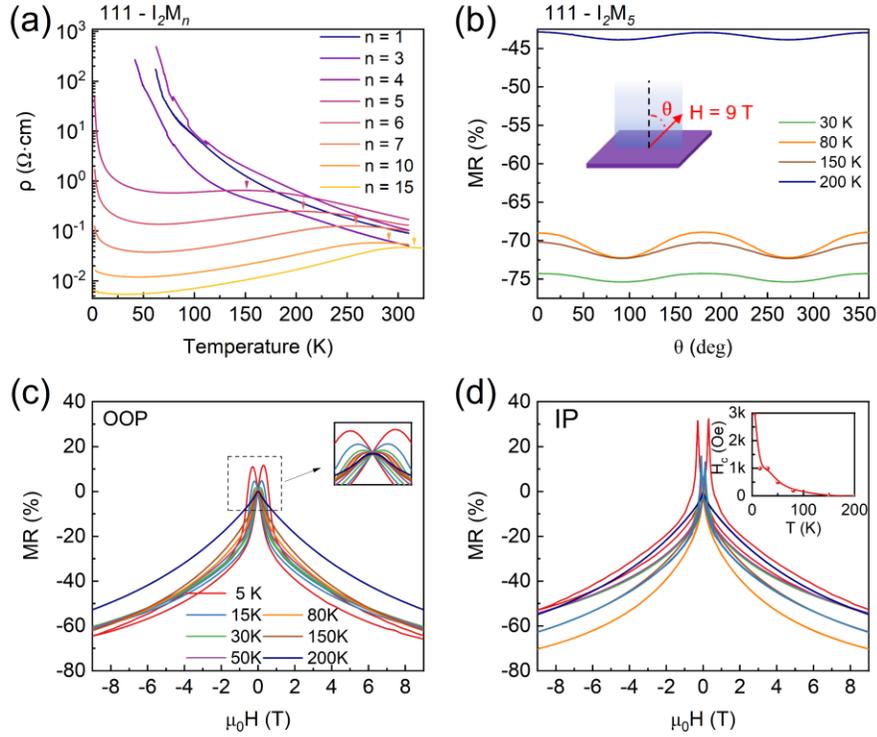

**FIG 2. Electrical transport properties of I$_2$M$_n$ superlattice samples.** (a) Temperature dependence of resistivity for LSMO with thickness *n* ranging from 1 to 15. The arrow indicates the metal-insulator transition temperature (T$_{MIT}$), which is extracted from the first-order differential of the R-T curve. (b) The angle dependence of the magnetoresistance under a 9 T field. The inset shows the configuration of the magnetic field. (c) The out-of-plane (OOP) magnetoresistance of the I$_2$M$_5$ superlattice sample on the external magnetic field. The inset is a local enlarged view near zero field. The FM loop appears below 150 K. (d) The angle dependence of the in-plane (IP) magnetoresistance of the I$_2$M$_5$ sample on the external magnetic field. The inset shows the temperature dependence of the coercivity H$_c$.

*Contact author: zliao@ustc.edu.cn
†Contact author: kaichen2021@ustc.edu.cn
#Contact author: ylgan@ustc.edu.cn
‡Contact author: siliang@nwu.edu.cn

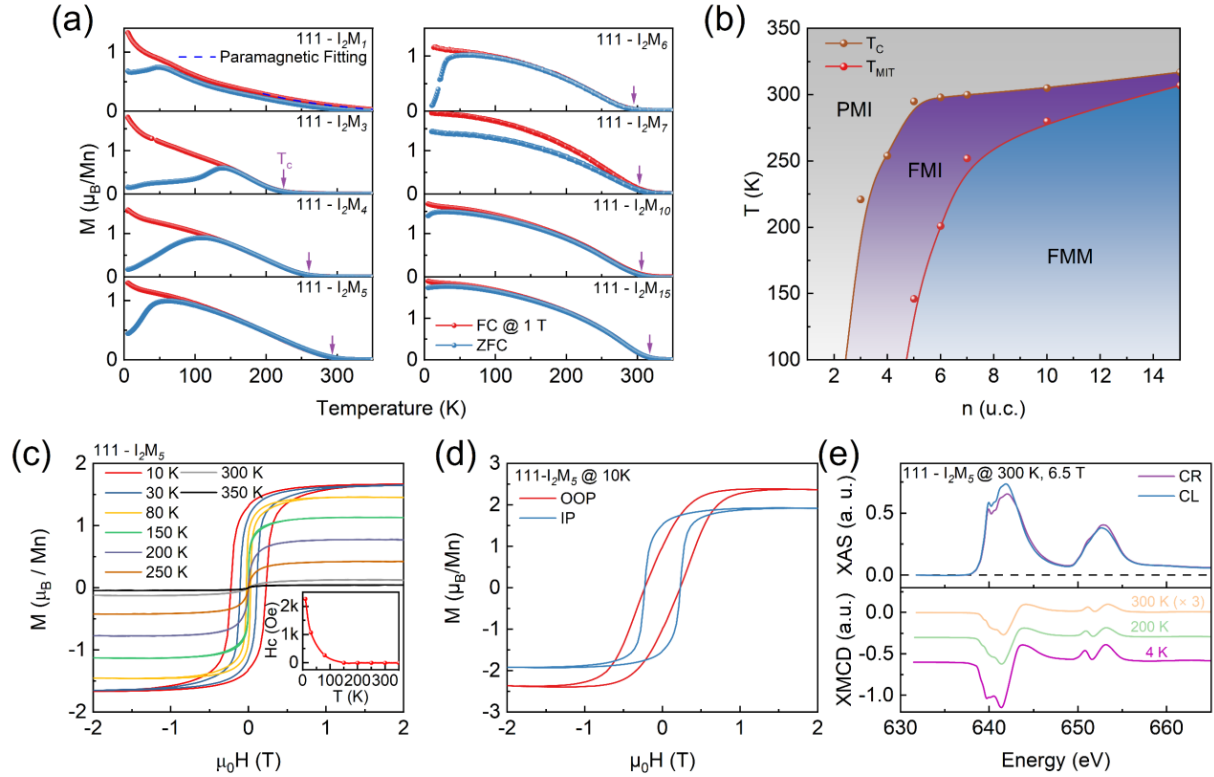

**FIG 3. Magnetic transport properties of I$_2$M$_n$ superlattice samples.** (a) Magnetic moment-temperature (MT) behavior of the I$_2$M$_n$ superlattice samples with a field cooling (FC) magnetic field of 1 T and a magnetic field of 0.1 T for both zero-field cooling (ZFC) and FC. (b) Phase diagram of superlattice samples. The gray area is the paramagnetic insulating phase (PMI), the purple area is the ferromagnetic insulating phase (FMI), and the blue area is the ferromagnetic metallic phase (FMM). The brown dot represents T$_c$, and the red dot represents T$_{MIT}$. (c) Temperature dependence of magnetization intensity-magnetic field loop of I$_2$M$_5$ superlattice sample. The inset shows the temperature dependence of the coercivity, which is consistent with the MR behavior. (d) Magnetic anisotropy behavior of I$_2$M$_5$ superlattice sample at 10 K. (e) XAS of Mn $L_{2,3}$-edge of I$_2$M$_5$ superlattice sample at room temperature and XMCD at 4 K, 200 K and 300 K. The XMCD signal of the I$_2$M$_5$ superlattice sample at 300 K is amplified three times.

*Contact author: zliao@ustc.edu.cn  
†Contact author: kaichen2021@ustc.edu.cn  
#Contact author: ylgan@ustc.edu.cn  
‡Contact author: siliang@nwu.edu.cn

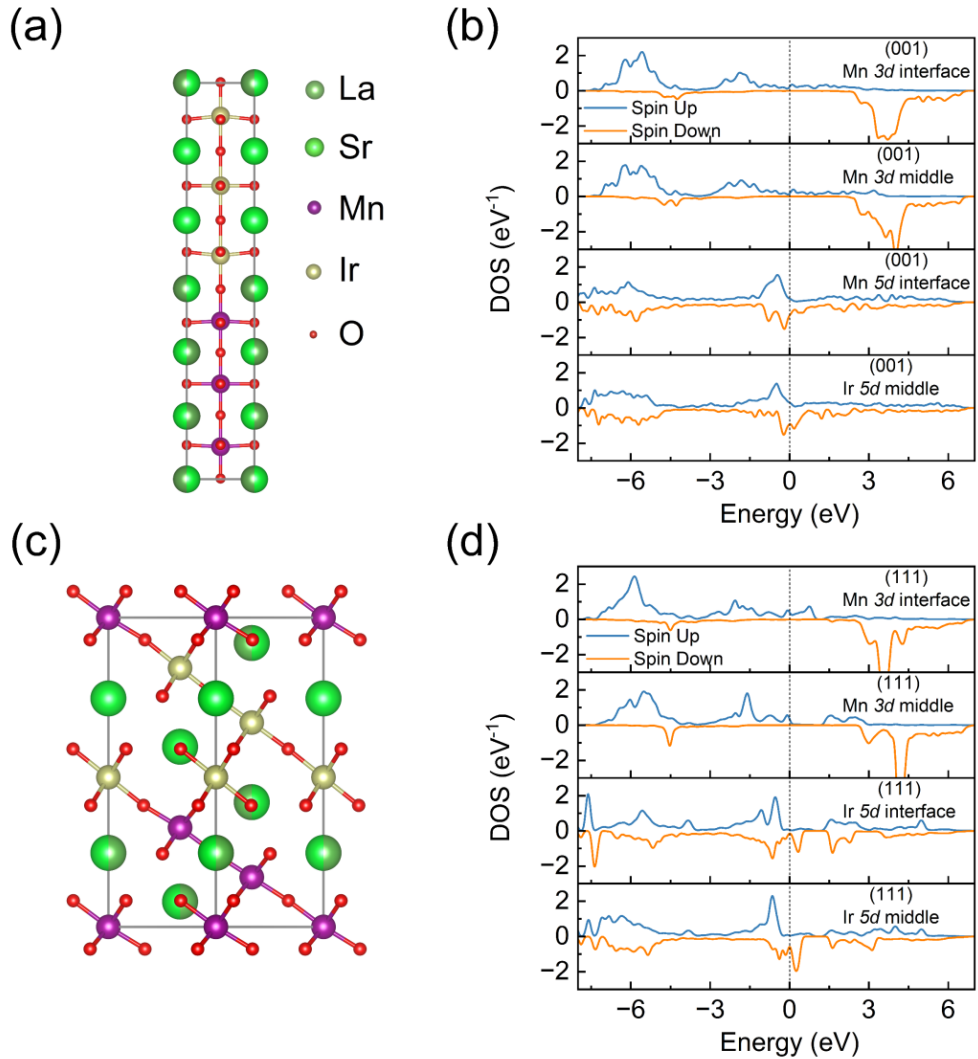

**FIG 4. DFT PDOS charge calculations.** Two heterostructures composed of LSMO and SIO along the (a) (001) and (c) (111) orientations, with the thicknesses of both the LSMO and SIO layers set to three u.c.. The orbital-resolved occupations and spin-resolved density of states Mn and Ir in the (b) (001)- and (d) (111)-oriented heterostructures.

*Contact author: zliao@ustc.edu.cn  
†Contact author: kaichen2021@ustc.edu.cn  
#Contact author: ylgan@ustc.edu.cn  
‡Contact author: siliang@nwu.edu.cn

# End Matter

In this End Matter, we provide a detailed WAL fitting analysis between $T_{MIT}$ and $T_c$. The significant positive MC (Fig. S10) generated by the CMR effect of LSMO can obscure the WAL. To clearly observe WAL, the CMR effect contribution must be subtracted [Fig. 5(a)], which can be accomplished by fitting using the Kohler rule. To minimize the influence of coercivity (or FM loop), the fitting is performed within the temperature range from $T_{MIT}$ to $T_c$. As shown in Fig. 5(b), by separating the MC signals into FM and interfacial SOC contributions, we find that the FM component contributes a typical WL behavior. In contrast, the MC attributed to interfacial SOC decreases to a cusp-like negative MC at the critical field ($\mu_0 H_c^{SOC}$), displaying characteristic WAL behavior. $\mu_0 H_c^{SOC}$ indicates the transition from WAL (when $\mu_0 H < \mu_0 H_c^{SOC}$) to WL (when $\mu_0 H > \mu_0 H_c^{SOC}$) induced by $H$, and it approximately reflects the strength of interface SOC, i.e., $H_{SO} \propto H_c^{SOC}$. At the same time, the WAL amplitude ($A_{WAL}$), defined as the absolute value of $\Delta\sigma_{xx}$ at $\mu_0 H_c^{SOC}$, reflects the interaction between interface SOC and WL, and is therefore proportional to $H_{so}/H_i$. At the same time, magnetic scattering in LSMO typically leads to WL, and it can be neglected in ultrathin LSMO [68,69]. Therefore, to quantify the interfacial SOC strength, the interfacial SOC contribution can be described by the two-dimensional localization theory of the diffusive region, following the Maekawa-Fukuyama formula [65] [Fig. 5(a)], while the FM-contributed CMR component can be fitted to the Kohler rule [Fig. 5(c)], thereby accurately describing the MC curves. Consequently, the total MC can be expressed as:

$$\frac{\Delta\sigma_{xx}(\mu_0 H)}{G_0} = -\psi\left(\frac{1}{2}+\frac{H_1}{H}\right) + \psi\left(\frac{1}{2}+\frac{H_2}{H}\right) + \frac{1}{2}\psi\left(\frac{1}{2}+\frac{H_3}{H}\right) - \frac{1}{2}\psi\left(\frac{1}{2}+\frac{H_i}{H}\right)$$
$$+ \left[-\ln\left(\frac{H_1}{H}\right) + \ln\left(\frac{H}{H_2}\right) + \frac{1}{2}\ln\left(\frac{H}{H_3}\right) - \frac{1}{2}\ln\left(\frac{H}{H_i}\right)\right]$$
$$+ \frac{\sigma_{xx}(0)}{G_0}\left(\frac{1}{A\cdot(\mu_0 H)^{\frac{1}{2}} + C\cdot(\mu_0 H)^{-\frac{1}{2}} + 1} - 1\right) \qquad (A1)$$

where $\psi(x)$ is the digamma function, the quantum of conductance $G_0 = \frac{e^2}{\pi h}$, and $H_1 = H_e + 2H_{SO}^x + H_{SO}^z = H_e + H_{SO}$, $H_2 = H_i + 2H_1\frac{H_{SO}^x + H_{SO}^z}{H_e - H_{SO}^z}$, $H_3 = H_i + 2H_1\frac{2H_{SO}^x}{H_e + H_{SO}^z - 2H_{SO}^x}$, where $H_e$, $H_i$, and $H_{so}$ are the effective fields related to the elastic, inelastic, and spin-orbit characteristic. The last Kohler rule is employed to describe the colossal magnetoresistance contributed by FM with fitting parameters A and C. As shown in the Fig. 5(d), optimal fits are achieved in the temperature range around $T_{MIT}$ to $T_c$.

Figure 5(c) illustrates that the MC contribution from the FM gradually diminishes with increasing temperature, with only a weak FM signal persisting at 300 K, consistent with the magnetic moment-temperature behavior discussed above. In contrast, the MC contribution from the interfacial SOC exhibits the opposite trend [Fig. 5(a)] and a typical WAL signal appears at 150 K. In most systems, WAL is observed only at low temperatures and gradually diminishes as the temperature increases. However, in our system, as ferromagnetism weakens (temperature increases), WAL displays anomalous behavior. The $\mu_0 H_c$ increases from ~0.15 T at 150 K to rare ~0.55 T at high temperatures (250 and 300 K) compared to other systems ($\mu_0 H_c \leq 0.2$ T) at lower temperature [65,70], indicating a strong SOC at the superlattice interface.

*Contact author: zliao@ustc.edu.cn
†Contact author: kaichen2021@ustc.edu.cn
#Contact author: ylgan@ustc.edu.cn
‡Contact author: siliang@nwu.edu.cn

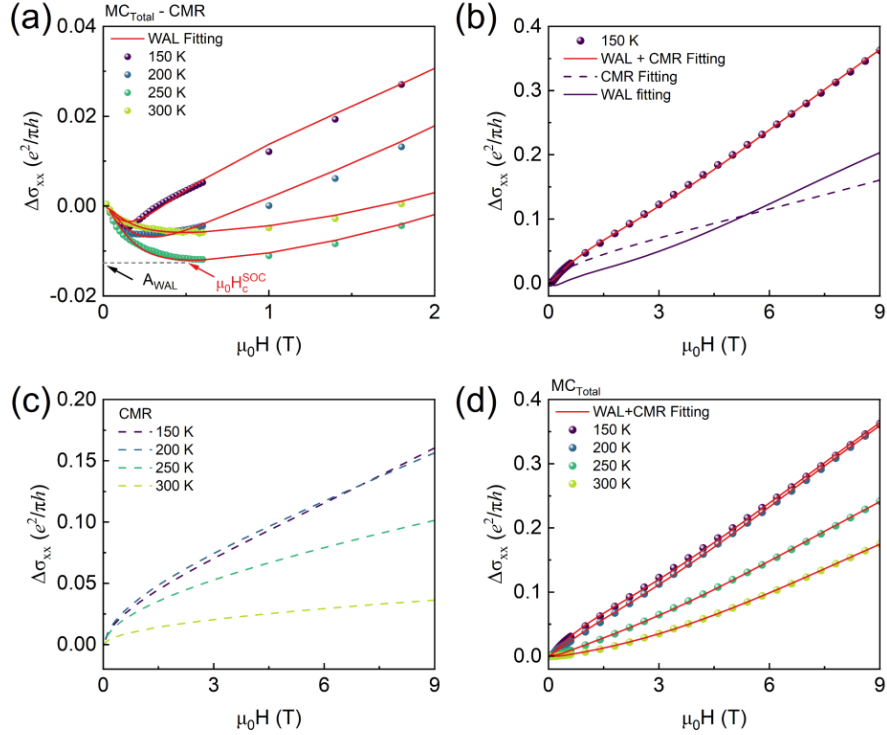

**FIG 5. Origin of Ferromagnetic insulating phase.** (a) Total MC minus the CMR contributions from 150 to 300 K. $A_{WAL}$ and $\mu_0 H_c^{SOC}$ represent phase coherent transmission and interface SOC, respectively. (b) Extracting the FM and interface SOC contributions from the MC at 150K. (c) Temperature dependence of extracting the CMR contributions from the MC. (d) Total MC fitting from 150 to 300 K.

*Contact author: zliao@ustc.edu.cn
†Contact author: kaichen2021@ustc.edu.cn
#Contact author: ylgan@ustc.edu.cn
‡Contact author: siliang@nwu.edu.cn

# Supplementary Materials for
# **Interfacial Spin-Orbit Coupling Induced Room Temperature Ferromagnetic Insulator**

Yuhao Hong,[1,2] Shilin Hu,[1] Ziyue Shen,[1] Chao Deng,[3] Xiaodong Zhang,[3] Lei Wang,[1] Long Wei,[1] Qinghua Zhang,[4] Lingfei Wang,[5] Liang Si,[3,6,7,‡] Yulin Gan,[1,#] Kai Chen,[1,†] Zhaoliang Liao[1,*]

*Contact author: zliao@ustc.edu.cn
†Contact author: kaichen2021@ustc.edu.cn
#Contact author: ylgan@ustc.edu.cn
‡Contact author: siliang@nwu.edu.cn

# I. Additional methods

### Sample growth
$(I_mM_n)_{10}$ superlattice samples are epitaxially grown on 111-oriented SrTiO$_3$ substrate by home-made pulsed laser deposition (PLD, KrF, $\lambda$ = 248 nm). The substrate is Ti-terminated by ultrasonic reacting in deionized water at 80 °C for 30 min and etching in BOE for 30 s. Then the STO substrate is annealed at 950 °C for 115 min to achieve an atomically smooth surface. The laser fluence during the growth process was 1.5 J/cm$^2$ for LSMO and 1.9 J/cm$^2$ for SIO, respectively. The typical substrate temperature was 775 °C and the oxygen partial pressure was 0.1 mbar. The thickness of each sublayer was precisely monitored by in situ RHEED during the growth, and then samples are gradually cooled down to room temperature in an oxygen partial pressure of 200 mbar.

### Structural characterization and STEM imaging
The X-ray reflectivity and reciprocal space mapping were characterized using an X-ray Diffractometer (Panalytical Empyrean Alpha 1). AFM images were performed on the Oxford Cypher (contact mode, AC240TSA-R3 tip). HAADF-STEM images were performed on an aberration-corrected FEI Titan Themis G2 operated at 300 kV at the Institute of Physics, Chinese Academy of science.

### Transport Measurement
All electrical transport measurements were performed using the Van der Pauw method in a PPMS (Quantum Design DynaCool system). Both in-plane and out-of-plane field-dependent magnetizations were measured by MPMS 3 (Quantum Design), and the actual sample magnetization was obtained by subtracting the linear diamagnetic contribution from the STO substrate.

### Synchrotron radiation experiments
The XMCD spectra for the Mn $L_{2,3}$-edge were carried out at Deimos beamline at synchrotron Soleil in total electron yield mode in normal incidence, at T = 4 K, 200 K and 300 K. The magnetic field of ± 6.5 Tesla, which is much higher than the saturation field, was applied along the beam during the XMCD measurements using left and right circularly polarized X-rays.

### DFT calculations
To study the charge transfer at the interface of (La,Sr)MnO$_3$ (LSMO) and SrIrO$_3$ (SIO), we constructed two 3:3 heterostructures with (001) and (111) orientations, where RHEED oscillations (Fig. S1d) reveal that the interfaces are IrO$_2$-SrO-MnO$_2$/SrO$_2$-Mn-LaSrO$_2$ and Ir-SrO$_3$-Mn/SrO$_3$-Mn-LaSrO$_3$. In each heterostructure, the thickness of both the LSMO and SIO layers was set to three perovskite unit cells.

Density-functional theory (DFT) [1,2] calculations were performed using the Vienna Ab-initio Simulation Package (VASP) [3-5] with the generalized-gradient approximation (GGA) in the Perdew-Burke-Ernzerhof (PBE) form [6] as the exchange-correlation functional. A plane-wave cutoff energy of 500 eV was used. The Brillouin zone was sampled with Monkhorst-Pack k-meshes of 10×10×1 for the (001) and 7×7×3 for the (111) heterostructures. All atoms were relaxed until the Hellmann-Feynman forces were below 0.01 eV/Å.

To address the role of electronic correlations, we performed calculations under four conditions: non-magnetic (NM), NM+Hubbard U, ferromagnetic+U (FM+U), and FM+U+SOC, with on-site Coulomb U applied to Mn-$3d$ and Ir-$5d$ orbitals following the Liechtenstein approach [7]. Spin-orbit coupling (SOC) was included in fully relativistic calculations. While we tested different initial magnetic moment configurations for Mn and Ir, the final self-consistent solutions were found to be independent of the initialization, converging to the same ground state electronic structure.

Bader charge analysis was carried out using the Henkelman group's code to evaluate charge transfer across the interface. In our PAW datasets, Mn was treated with a $d^6s^1$ valence (7 electrons) and Ir with $d^8s^1$ (9 electrons), so the reported negative Bader charges indicate the number of electrons lost relative to the pseudopotential valence. Electronic structure and band calculations were further checked with the WIEN2k code [8]. Maximally localized Wannier functions (MLWFs) [9] were obtained using the Wien2Wannier [10] interface and Wannier90 [11] to project Mn-$3d$ and Ir-$5d$ bands, allowing us to extract orbital-resolved densities of states and occupation numbers for Mn and Ir.

*Contact author: zliao@ustc.edu.cn  
†Contact author: kaichen2021@ustc.edu.cn  
#Contact author: ylgan@ustc.edu.cn  
‡Contact author: siliang@nwu.edu.cn

## II. Growth of Superlattice films

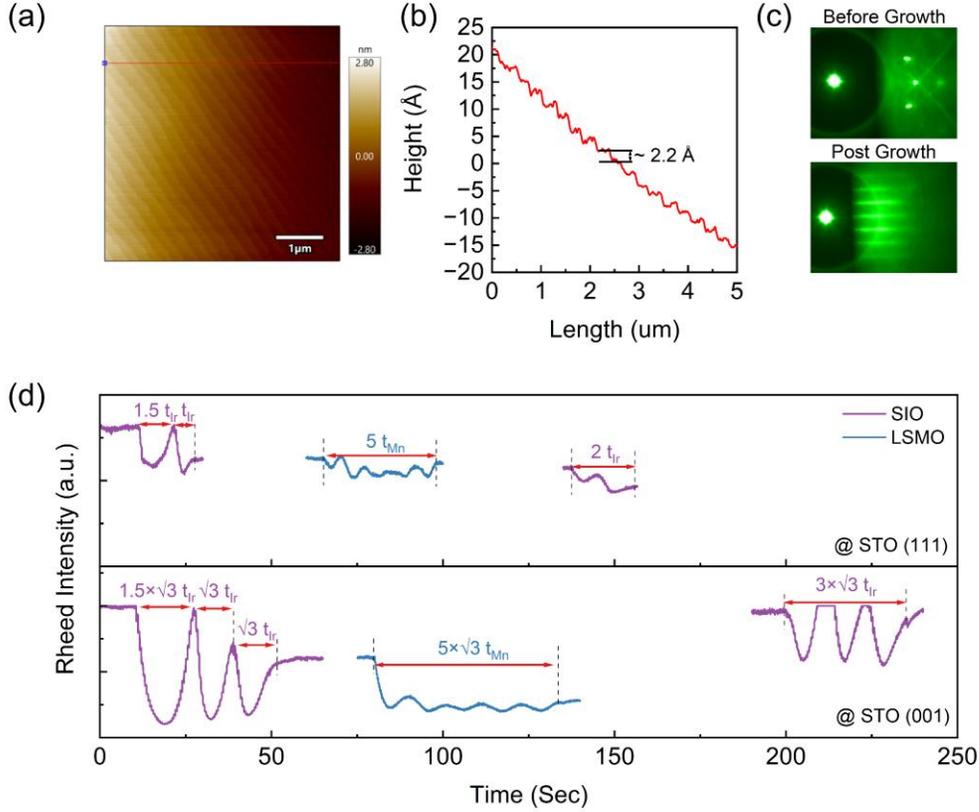

FIG S1. Surface and growth characterization. (a) AFM surface topography of atomically smooth Ti-terminated STO (111) substrate. (b) Measured step heights of STO (111) substrate from the red line in (a). (c) RHEED patterns before and post growth. (d) Comparison of RHEED intensity oscillations during the growth of SIO and LSMO on STO (001) and (111) substrates. The upper figure shows the RHEED oscillations of the film grown on STO (111) substrate, and the lower figure shows the RHEED oscillations of the film grown on STO (001) substrate. The purple curve represents the growth of SIO film, and the blue curve represents the growth of LSMO film.

The highly smooth AFM surface morphology [Fig. S1(a)] indicates that an atomically smooth Ti-terminated STO (111) substrate has been achieved, with measured step heights [Fig. S1(b)] consistent with the crystal structure. The RHEED patterns before and after growth [Fig. S1(c)] both exhibited well-defined two-dimensional structures. As Fig. S1(d) shows that during the initial stages of growth, RHEED monitoring consistently showed a clear layer-by-layer growth mode. The time required to grow 1 u.c. of SIO and LSMO on the STO (111) substrate is defined as $t_{Ir}$ and $t_{Mn}$, respectively, while the time required to grow 1 u.c. on the STO (001) substrate is approximately $\sqrt{3}$ times that on the STO (111). Approximating the perovskite phases of LSMO and SIO as cubic systems, the required growth time perfectly aligns with the Bragg formula: $d_{111} = \frac{d_{001}}{\sqrt{1^2+1^2+1^2}} = \frac{d_{001}}{\sqrt{3}}$. Where d is the interplanar spacing.

Meanwhile, the time required to grow the first u.c. of SIO is approximately 1.5 times longer than that for each subsequent u.c., as observed on both STO (001) and STO (111) substrates, which is due to the decomposition of $IrO_6$ octahedrons to form SrO and $SrO_3$ terminated surfaces [12]. According to the report of Xuan Zheng et al. [13], the growth time per RHEED intensity oscillation for monoclinic SIO on STO (111) substrates is three times longer than that for perovskite SIO. In our study, the time for each RHEED oscillation intensity consistent with the characteristics of the perovskite phase. This indicates that the in-plane compressive stress can effectively maintain the perovskite phase of the SIO thin film.

*Contact author: zliao@ustc.edu.cn
†Contact author: kaichen2021@ustc.edu.cn
#Contact author: ylgan@ustc.edu.cn
‡Contact author: siliang@nwu.edu.cn

## III. HAADF-STEM image and ELLS mapping images

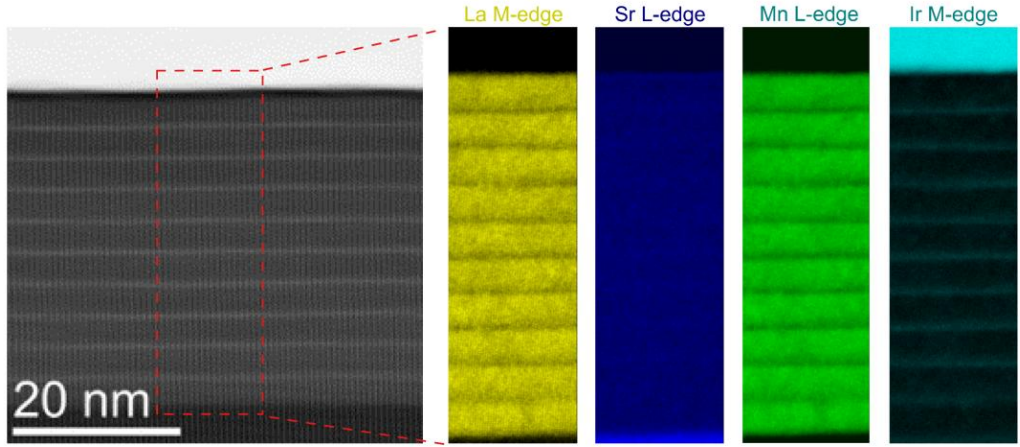

FIG S2. HAADF-STEM image of the $I_2M_{15}$ superlattice sample and EELS mapping image of the red frame area. The yellow is the element distribution of La, the purple is the element distribution of Sr, the green is the element distribution of Mn, and the blue is the element distribution of Ir.

The HAADF-STEM and ELLS images of La, Mn, and Ir all reveal clear superlattice interfaces, with no significant elemental diffusion observed in the EELS. It is worth noting that the $M_4$ edge of Ir (~2116 eV) and the $M_5$ edge of Pt (~2122 eV) are very close. As a result, the Pt protective layer on the sample surface also appears in the Ir ELLS images. Moreover, only two layers of Ir atoms were observed per superlattice period [Fig. 1(c) in main manuscript], in contrast to the three layers of Ir atoms per u.c. in monoclinic SIO. This observation further rules out the possibility of a monoclinic SIO phase.

*Contact author: zliao@ustc.edu.cn
†Contact author: kaichen2021@ustc.edu.cn
#Contact author: ylgan@ustc.edu.cn
‡Contact author: siliang@nwu.edu.cn

## IV. Structural properties of I$_2$M$_n$ superlattice Samples

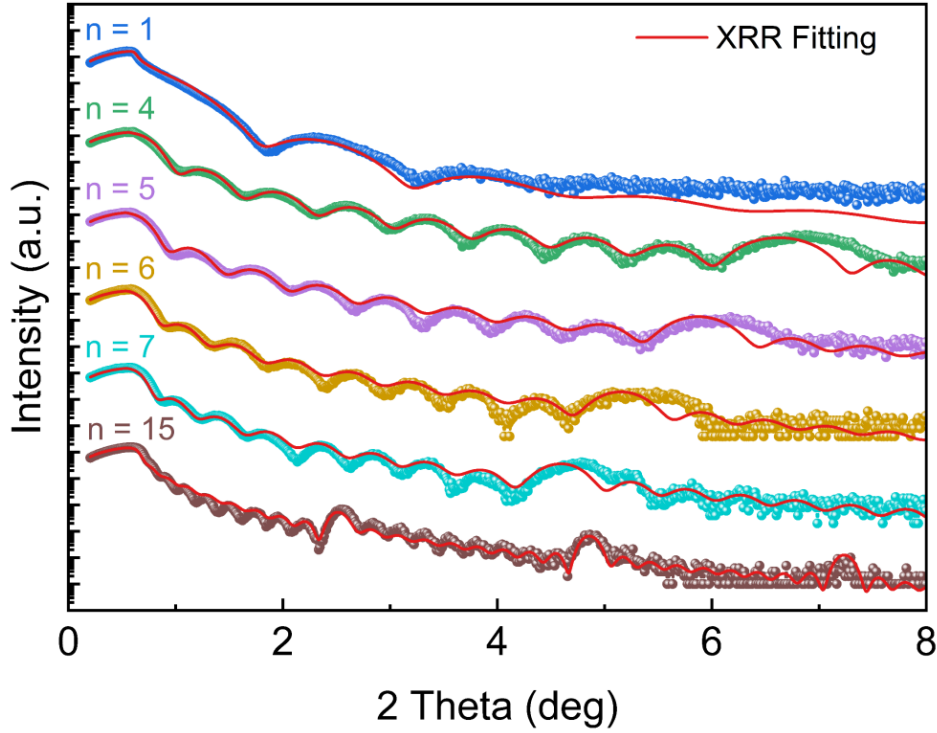

FIG S3. XRR of I$_2$M$_n$ superlattice samples and GenX fitting. The dots are the original XRR data, from top to bottom they are n = 1, 4, 5, 6, 7, and 15. The red curve is the result of fitting by GenX software.

As n increases, the superlattice peaks in the XRR gradually shift towards the low-angle region, aligning with expectations. Moreover, the superlattice structure fitted using GenX software closely matches the original data, further indicating that high-quality superlattice samples have been obtained.

*Contact author: zliao@ustc.edu.cn
†Contact author: kaichen2021@ustc.edu.cn
#Contact author: ylgan@ustc.edu.cn
‡Contact author: siliang@nwu.edu.cn

## V. RSM images of I$_2$M$_n$ superlattice samples

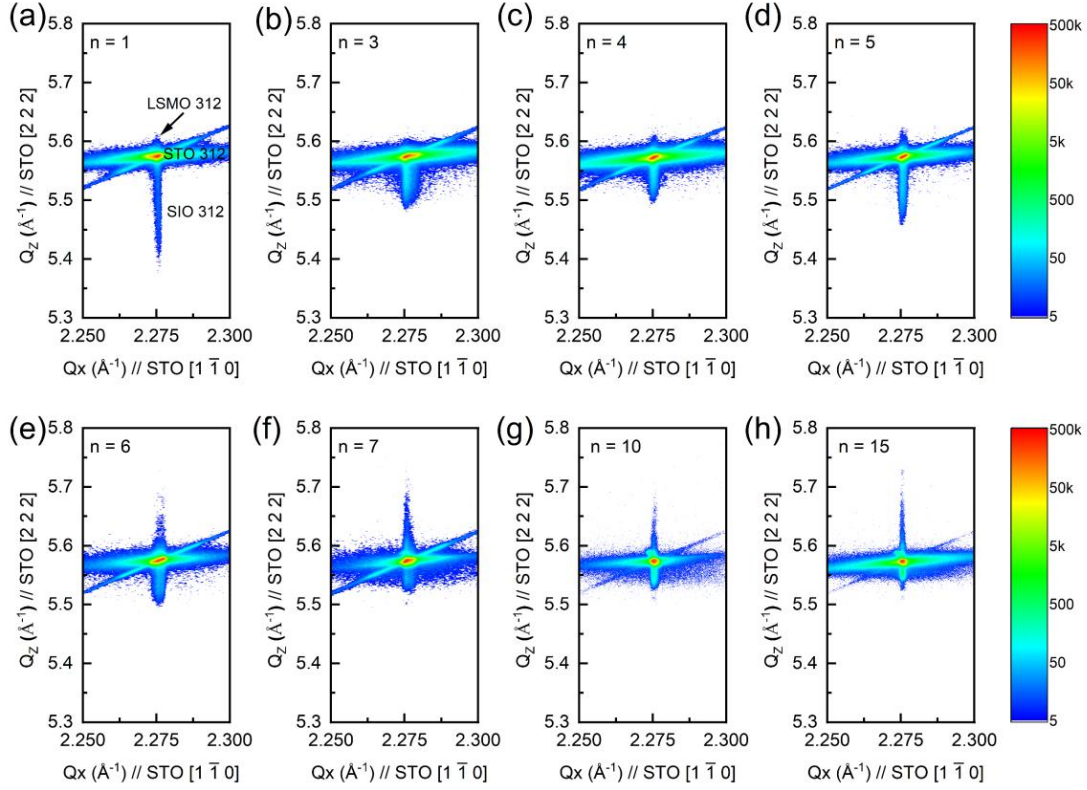

FIG S4. RSM images of I$_2$M$_n$ superlattice samples. SIO (lower) and LSMO (upper) peaks surrounding the STO (312) peak, and n from 1 to 15. All images are color scaled consistently

The RSM reveals that the in-plane lattice constants of the LSMO and SIO thin films in all samples are well-matched with those of the substrate, indicating excellent epitaxial quality. As n increases, the out-of-plane relaxation of the SIO thin film gradually decreases, while that of the LSMO increases.

*Contact author: zliao@ustc.edu.cn  
†Contact author: kaichen2021@ustc.edu.cn  
#Contact author: ylgan@ustc.edu.cn  
‡Contact author: siliang@nwu.edu.cn

## VI. Magnetic transport properties of $I_mM_5$ and $T_2M_5$ superlattice samples

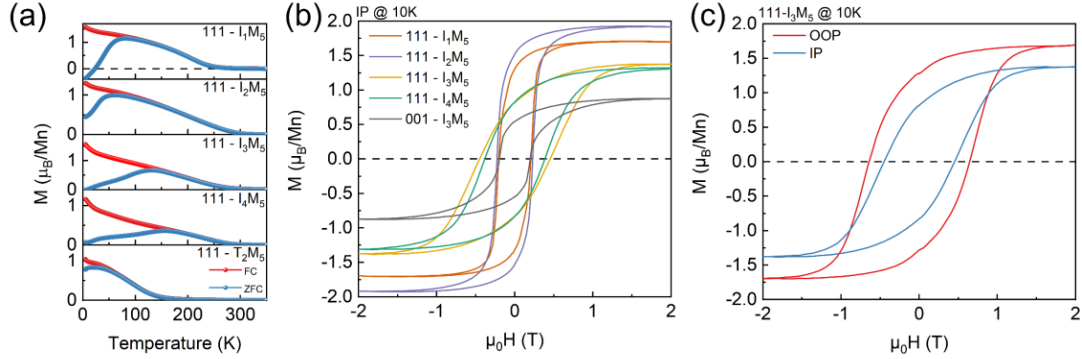

FIG S5. Effect of SIO thickness and substrate orientation on magnetic transport in superlattice samples. a, Dependence between MT curve and SIO. b, Dependence of magnetic anisotropy on SIO and substrate orientation at 10 K. c, Out-of-plane (red line) and in-plane (blue line) hysteresis loops of $I_3M_5$ superlattice samples at 10 K.

As the thickness m of the SIO in the $I_mM_5$ superlattice samples increases, no significant change in the $T_c$ is observed in Fig. S5(a). However, when SIO is replaced with STO, the $T_c$ immediately decreases to 150K, indicating that the increase in $T_c$ is induced by SIO. As shown in Fig. S5(b), the saturation magnetization of the samples does not increase monotonically with m; instead, it reaches a maximum at m = 2, suggesting that the SIO film itself does not exhibit ferromagnetism. Additionally, the saturation magnetization of the (111)-oriented superlattice samples is greater than that of the (001)-oriented samples, indicating that the (111) orientation further enhances the ferromagnetism of the samples. Moreover, when m ≥ 3, changes in the magnetic anisotropy of the samples are observed. Fig. S5(c) further confirms that the magnetic easy axis of the $I_3M_5$ sample tends to be out-of-plane.

*Contact author: zliao@ustc.edu.cn  
†Contact author: kaichen2021@ustc.edu.cn  
#Contact author: ylgan@ustc.edu.cn  
‡Contact author: siliang@nwu.edu.cn

# VII. Variable temperature XMCD

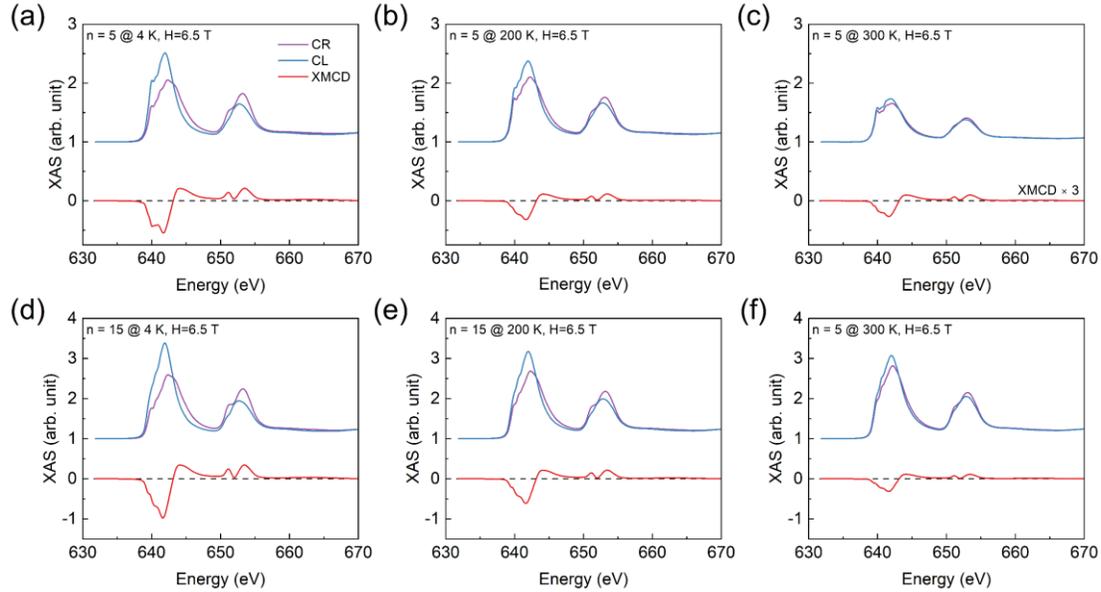

FIG S6. Temperature-dependent XAS and XMCD spectra of $L_{2,3}$ edges of $I_2M_5$ and $I_2M_{15}$ superlattice samples. All testes were under an external field of 6.5 T, with the incident X-ray angle of 35 degrees to the sample surface, using circular left (CL, blue line) and right (CR, purple) polarized X-rays. The XMCD spectrum (red line) is obtained by subtracting the CR spectrum from the CL spectrum. The XMCD signal of the $I_2M_5$ superlattice sample at 300 K was amplified three times

Both $I_2M_5$ and $I_2M_{15}$ samples exhibit similar XMCD behavior, with a non-zero XMCD signal observable at 300 K. In the $I_2M_5$ sample, LSMO experiences strong in-plane tensile strain from the substrate and SIO, which has not relaxed, as confirmed by Fig. S4(d). This results in an absorption peak at ~640 eV [Fig. S6(a-c)]. In contrast, for the $I_2M_{15}$ sample, Fig. S4(h) shows that the in-plane tensile strain in the LSMO has relaxed, and thus no absorption peak is observed at ~640 eV [Fig. S6(d-h)].

*Contact author: zliao@ustc.edu.cn
†Contact author: kaichen2021@ustc.edu.cn
#Contact author: ylgan@ustc.edu.cn
‡Contact author: siliang@nwu.edu.cn

## VIII. Oxygen octahedron tilt imaging

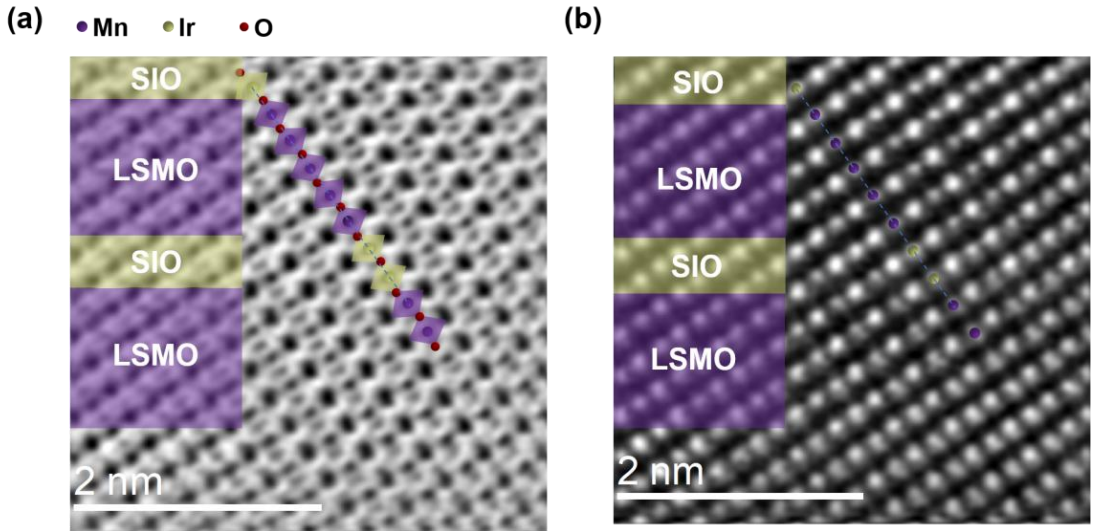

FIG S7. ABF-STEM (a) and HAADF-STEM (b) image of the $I_2M_5$ superlattice sample.

Both LSMO and SIO retain oxygen octahedral rotations. Notably, the LSMO layer thickness in each superlattice period (5 u.c.) is well below the critical thickness for the proximity effect (8 u.c.) [14]. This demonstrates that the proximity effect is not the origin of the room-temperature ferromagnetic insulating phase observed in our work.

*Contact author: zliao@ustc.edu.cn  
†Contact author: kaichen2021@ustc.edu.cn  
#Contact author: ylgan@ustc.edu.cn  
‡Contact author: siliang@nwu.edu.cn

## IX. Site-resolved simplified DFT calculations

| 001 | Mn interface | Mn middle | Mn average | Ir interface | Ir middle | Ir average | Mn bulk | Ir bulk |
|---|---|---|---|---|---|---|---|---|
| **Total** | 3.892 | 3.658 | 3.814 | 4.992 | 4.573 | 4.852 | 3.667 | 5.000 |
| $d_{xy}$ | 1.608 | 1.375 | 1.530 | 1.467 | 1.539 | 1.492 | 1.222 | 1.667 |
| $d_{yz}/d_{xz}$ | 1.123 | 1.138 | 1.128 | 1.672 | 1.491 | 1.612 | 1.222 | 1.667 |
| **111** | **Mn interface** | **Mn middle** | **Mn average** | **Ir interface** | **Ir middle** | **Ir average** | **Mn bulk** | **Ir bulk** |
| **Total** | 3.694 | 3.582 | 3.657 | 5.022 | 4.981 | 5.008 | 3.667 | 5.000 |
| $d_{xy}$ | 1.198 | 1.175 | 1.190 | 1.645 | 1.625 | 1.638 | 1.222 | 1.667 |
| $d_{yz}/d_{xz}$ | 1.200 | 1.176 | 1.190 | 1.645 | 1.625 | 1.639 | 1.222 | 1.667 |

Table S1. Orbital-resolved electron occupation from (non-magnetic) DFT calculations. As a comparison, orbital occupations of nominal bulk LSMO and SIO are also shown in the last two columns.

As presented in Table S1, our results reveal that at the (001)-oriented interface, the charge transfer across the SrO layer amounts to 3.814-3.667=0.147 electrons. Here, 3.814 (electron/Mn) represents the average *3d*-orbital occupation for Mn in the (001)-heterostructure, while 3.667 is the corresponding value for Mn in bulk LSMO. This charge transfer occurs from the Ir-5d orbitals to the Mn-*3d* orbitals. In contrast, the charge transfer at the (111)-oriented interface is significantly smaller, at 3.657-3.667= -0.010 electrons, where 3.657 is the average Mn occupation in the (111)-heterostructure. This effectively indicates negligible charge transfer at the (111)-interface.

*Contact author: zliao@ustc.edu.cn  
†Contact author: kaichen2021@ustc.edu.cn  
#Contact author: ylgan@ustc.edu.cn  
‡Contact author: siliang@nwu.edu.cn

# X. SOC effect on electron occupation

(a) (001)-Oriented (SIO)$_3$/(LSMO)$_3$

(b) (111)-Oriented (SIO)$_3$/(LSMO)$_3$

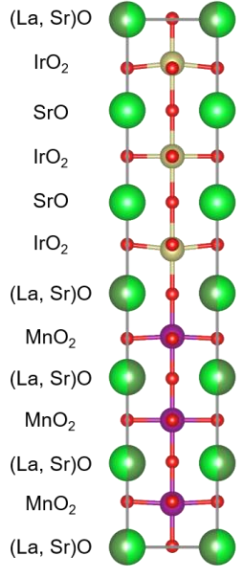
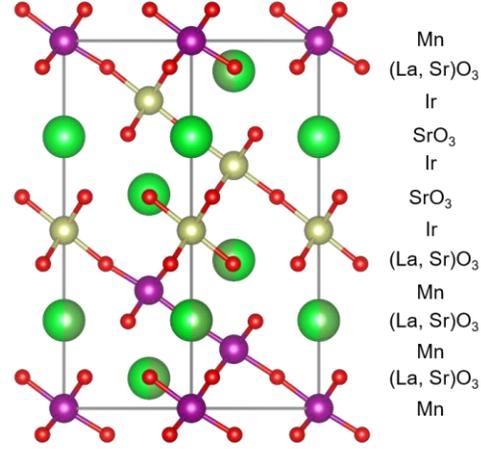

FIG S8. Employed crystal structures of LSMO:SIO (3:3) superlattice along (001) and (111), respectively.

To study the effect of SOC, we performed calculations with different setups: nonmagnetic (NM), nonmagnetic + Hubbard U (NM+U), ferromagnetic + Hubbard U (FM+U), and ferromagnetic + Hubbard U + spin-orbit coupling (FM+U+SOC). The thicknesses of the LSMO layer and the SIO layer were set as three, respectively. The doped Sr in LaO and LaO$_3$ layers was achieved by employing Virtual Crystal Approximation (VCA).

| *Atom* | *Mn* | *Ir* | *La* | *Sr* | *O* | *Mn* | *Ir* | *La* | *Sr* | *O* |
|---|---|---|---|---|---|---|---|---|---|---|
| ***Orientation*** | **(111)** | | | | | **(001)** | | | | |
| ***NM*** | -1.546 | -1.498 | -1.339 | -2.607 | 1.095 | -1.657 | -1.644 | -1.331 | -2.610 | 1.136 |
| ***NM+U*** | -1.584 | -1.447 | -1.342 | -1.615 | 1.094 | -1.662 | -1.593 | -2.330 | -2.613 | 1.128 |
| ***FM+U*** | -1.684 | -1.473 | -1.335 | -2.608 | 1.113 | -1.692 | -1.613 | -1.332 | -2.612 | 1.137 |
| ***FM+SOC+U*** | -1.685 | -1.541 | -1.336 | -2.605 | 1.124 | -1.690 | -1.651 | -1.330 | -2.609 | 1.142 |

Table S2 Average number of charge transfer in different crystal orientation superlattices obtained from Bader charge calculations.

In our DFT Bader charge calculations, the PAW pseudopotentials include 7 valence electrons for Mn ($d^6s^1$) and 9 valence electrons for Ir ($d^8s^1$). Here, the negative value (resulted by Bader charge minus valence electrons) indicates a loss of electrons relative to the pseudopotential valence electron count. For Mn and Ir cations, their valence states and electron numbers deviate from their bulk nominal valence states: +3.3 for Mn and +4 for Ir, and $3d^{3.67}$ in Mn and $5d^5$ in Ir, respectively. This indicates that in the (001) and (111) crystal orientation heterostructure, Mn and Ir both form strong covalent bonds with O ions.

*Contact author: zliao@ustc.edu.cn
†Contact author: kaichen2021@ustc.edu.cn
#Contact author: ylgan@ustc.edu.cn
‡Contact author: siliang@nwu.edu.cn

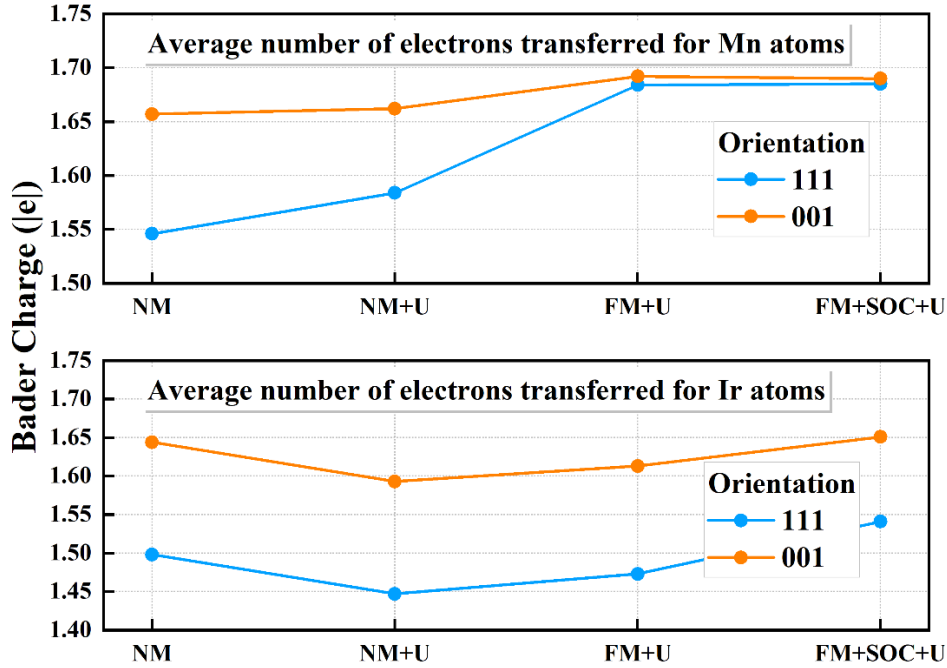

FIG S9. Average number of charge transfer per Mn and Ir atom in (111) and (001) crystal orientation superlattices obtained from Bader charge calculations.

For the FM+SOC+U configuration, the estimated valence states and number of the rest electron in Mn-*3d* and Ir-*5d* orbitals are:
(001):
Mn(+1.690): 7 - 1.690 → +5.31e
Ir(+1.651): 9 - 1.651 → +7.35e
(111):
Mn(+1.658): 7 - 1.658 → +5.34e
Ir(+1.541): 9 - 1.541 → +7.46e
In summary, in the DFT Bader charge results: (001) shows a slightly larger electron loss for both Mn and Ir compared to (111), implying stronger covalency in Mn-O and Ir-O bonds in the (001) orientation due to enhanced Mn-O and Ir-O hybridization. For instance, in (001), Ir deviates further from the nominal $Ir^{4+}$ ($5d^5$) state in bulk SIO, consistent with stronger Ir-O covalent bonding at the interface. In (111), Mn and Ir have slightly lower oxidation states (fewer electrons lost), suggesting that neighboring oxygen atoms must carry a small hole population, while Mn and Ir *d*-orbitals are slightly more occupied by electrons, being similar to their bulk states.
Finally, in FM+SOC+U calculations for (001) orientation, Mn and Ir have similar valences while in (111), the Ir has low valence state and hence 2.3 electrons more occupation. As a consequence, the above calculations demonstrate that in (001) orientation, the Ir occupies roughly two more electrons than Mn, while in (111) orientation, the Ir occupies 2.1 electrons more than Mn, hence in (111) orientation there are less electrons transfer from Ir to Mn. These findings quantitatively support our previous DFT-Wannier occupation calculations in our last version of submission and the original statement that the (111) orientation exhibits negligible interfacial charge transfer compared to the (001) orientation.

*Contact author: zliao@ustc.edu.cn
†Contact author: kaichen2021@ustc.edu.cn
#Contact author: ylgan@ustc.edu.cn
‡Contact author: siliang@nwu.edu.cn

## XI. SOC effect on magnetism

| (001) | Mn(μB) | Ir(μB) | Total(μB) | (111) | Mn(μB) | Ir(μB) | Total(μB) |
|---|---|---|---|---|---|---|---|
| **FM+U** | 3.55 | 0.52 | 12.86 | **FM+U** | 3.63 | 0.67 | 13.84 |
| **FM+U+SOC** | 3.58 | 0.06 | 10.31 | **FM+U+SOC** | 3.63 | 0.32 | 11.50 |

**Table. S3** *DFT+U and DFT+U+SOC magnetic moment projected onto Mn, Ir and total moment for (001) and (111) orientation LSMO:SIO superlattice.*

To examine the impact of SOC on magnetism, we computed the projected magnetic moments on Ir and Mn for both FM+U and FM+U+SOC cases in the (001) and (111) orientation superlattices. The results are summarized in the **Table S3**. In the FM+U calculations, Ir exhibits a magnetic moment of 0.5-0.7 $\mu_B$ per Ir atom for both the (001) and (111) superlattices. After including SOC, the Ir moment in the (001) superlattice is fully quenched to <0.1$\mu_B$, whereas in the (111) superlattice it is only partially reduced and remains above 0.3 $\mu_B$ per Ir. These results indicate that in the (111) superlattice, the influence of SOC on Ir persists.

*Contact author: zliao@ustc.edu.cn
†Contact author: kaichen2021@ustc.edu.cn
#Contact author: ylgan@ustc.edu.cn
‡Contact author: siliang@nwu.edu.cn

## XII. MC fitting at variable temperature

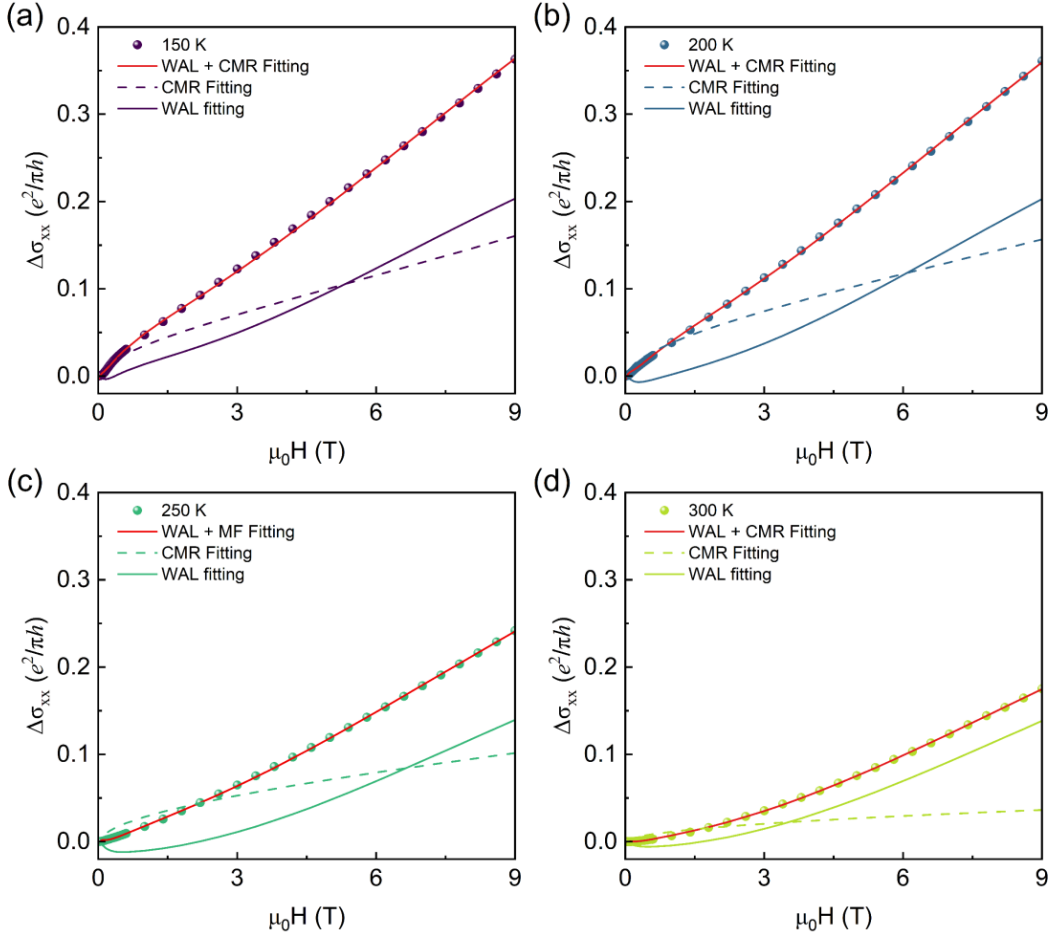

FIG S10. Temperature-dependent MC fitting of $I_2M_5$ superlattice from 150 K to 300 K, from which the contributions of SOC and MF are extracted. The dots are the original MC data, the red curve is the total MC fitting data, the colored solid line is the extracted SOC contribution, and the colored dotted line is the extracted CMR contribution.

The magnetoconductance is composed of two factors: the colossal magnetoresistance effect contributed by LSMO ferromagnetism and weak anti-localization. The colossal magnetoresistance effect can be expressed using the empirical formula - Kohler's Rule, as follows:

$$\text{MR} = \alpha \left(\frac{\mu_0 H}{\rho_0}\right)^{1/2} + \beta \left(\frac{\mu_0 H}{\rho_0}\right)^{-1/2}$$

Where $\alpha$ and $\beta$ are the coefficients of the fitting, and $\rho_0$ is the resistivity at zero field.

In a diffusive regime, the weak anti-localization caused by SOC can be described by the Maekawa-Fukuyama formula, neglecting Zeeman splitting, as follows:

$$\frac{\Delta\sigma_{xx}(\mu_0 H)}{G_0} = -\psi\left(\frac{1}{2} + \frac{\mu_0 H_1}{\mu_0 H}\right) + \psi\left(\frac{1}{2} + \frac{\mu_0 H_2}{\mu_0 H}\right) + \frac{1}{2}\psi\left(\frac{1}{2} + \frac{\mu_0 H_3}{\mu_0 H}\right) - \frac{1}{2}\psi\left(\frac{1}{2} + \frac{\mu_0 H_i}{\mu_0 H}\right)$$
$$+ \left[-\ln\left(\frac{\mu_0 H_1}{\mu_0 H}\right) + \ln\left(\frac{\mu_0 H}{\mu_0 H_2}\right) + \frac{1}{2}\ln\left(\frac{\mu_0 H}{\mu_0 H_3}\right) - \frac{1}{2}\ln\left(\frac{\mu_0 H}{\mu_0 H_i}\right)\right]$$

*Contact author: zliao@ustc.edu.cn  
†Contact author: kaichen2021@ustc.edu.cn  
#Contact author: ylgan@ustc.edu.cn  
‡Contact author: siliang@nwu.edu.cn

Where ψ(x) is the digamma function, the quantum of conductance $G_0 = \frac{e^2}{\pi h}$, and $H_1 = H_e + 2H_{SO}^x + H_{SO}^z = H_e + H_{SO}$, $H_2 = H_i + 2H_1 \frac{B_{SO}^x + B_{SO}^z}{B_e - B_{SO}^z}$, $H_3 = H_i + 2H_1 \frac{2H_{SO}^x}{H_e + H_{SO}^z - 2H_{SO}^x}$, where $H_e$, $H_i$ and $H_{so}$ are the effective fields related to the elastic, inelastic and spin-orbit characteristic..

Therefore, the total magnetoconductance $MC_{total}$ can be expressed as:

$$MC_{total}(\mu_0 H) = -\psi\left(\frac{1}{2} + \frac{H_1}{H}\right) + \psi\left(\frac{1}{2} + \frac{H_2}{H}\right) + \frac{1}{2}\psi\left(\frac{1}{2} + \frac{H_3}{H}\right) - \frac{1}{2}\psi\left(\frac{1}{2} + \frac{H_i}{H}\right)$$
$$+ \left[-\ln\left(\frac{H_1}{H}\right) + \ln\left(\frac{H}{H_2}\right) + \frac{1}{2}\ln\left(\frac{H}{H_3}\right) - \frac{1}{2}\ln\left(\frac{H}{H_i}\right)\right]$$
$$+ \frac{\sigma_{xx}(0)}{G_0}\left(\frac{1}{A(\mu_0 H)^{1/2} + C(\mu_0 H)^{-1/2} + 1} - 1\right)$$

Where A and C are the coefficients of the fitting.

*Contact author: zliao@ustc.edu.cn
†Contact author: kaichen2021@ustc.edu.cn
#Contact author: ylgan@ustc.edu.cn
‡Contact author: siliang@nwu.edu.cn

## XIII. The typical length

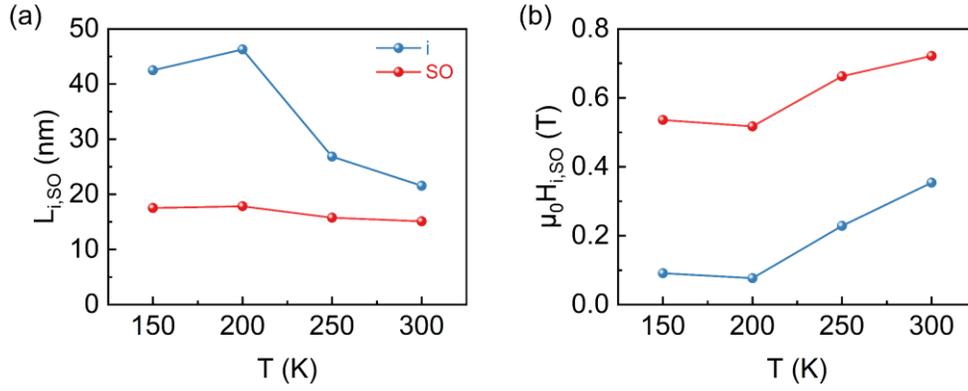

FIG S11. Feature quantities extracted from MC. (a) Phase coherence length ($L_i$, blue line) and spin relaxation length ($L_{SO}$, red line) as a function of temperature. (b) Effective fields related to the inelastic ($\mu_0 H_i$, blue line) and spin-orbit characteristic ($\mu_0 H_{SO}$, red line) as a function of temperature.

Here the typical phase coherence length ($L_i$) and spin relaxation length ($L_{SO}$) have been derived from the relation $L_{i,SO} = \sqrt{\hbar/(4eB\mu_0 H_{i,SO})}$

*Contact author: zliao@ustc.edu.cn
†Contact author: kaichen2021@ustc.edu.cn
#Contact author: ylgan@ustc.edu.cn
‡Contact author: siliang@nwu.edu.cn

*Contact author: zliao@ustc.edu.cn
†Contact author: kaichen2021@ustc.edu.cn
#Contact author: ylgan@ustc.edu.cn
‡Contact author: siliang@nwu.edu.cn